\documentclass[12pt]{article}
\let\section=\subsection     \let\subsection=\subsubsection                
\usepackage{epsfig}
\begin{document}
%
\begin{center}
   {\large \bf Quasiparticles in QCD thermodynamics}\\[2mm]
   {\large \bf and applications\footnote{Work supported in part by BMBF and GSI.}}\\[5mm]
   R.~A.~Schneider$^a$, T.~Renk$^a$ and W.~Weise$^{ab}$ \\[5mm]
   {\small \it  $^a$ Physik-Department, Technische Universit\"at M\"unchen,\\
   85747 Garching/M\"unchen, Germany \\[2mm]
   $^b$ECT*, Villa Tambosi, 38050 Villazzano (Trento), Italy \\[8mm]}
\end{center}
%
%
\begin{abstract}\noindent
We propose a novel quasiparticle interpretation of the equation of state of deconfined QCD at finite
temperature. Using appropriate thermal masses, we introduce a phenomenological parametrization of the
onset of confinement in the vicinity of the phase transition. Lattice results of the energy
density, the pressure and the interaction measure of pure SU(3) gauge theory are well reproduced. A relation between the thermal energy density  of the Yang-Mills vacuum and the
chromomagnetic condensate $\langle \mathbf{B}^2 \rangle_T$ is found. We also
present the two flavour QCD equation of state for realistic quark masses and apply the model to dilepton production in ultrarelativistic heavy-ion collisions.
\end{abstract}
%
%
%
\section{Introduction}
%
QCD undergoes a transition from a confined hadronic phase to a deconfined partonic phase, the
quark-gluon plasma (QGP), at a temperature of $T_c \sim 150 - 170$ MeV \cite{LAT1}. A central quantity of matter in thermal equilibrium is the Helmholtz free energy from
which the pressure $p$, energy density $\epsilon$ and entropy density $s$ -- which are important
ingredients for the description of ultra-relativistic heavy-ion collisions -- can be derived. A
first-principles understanding of the equation of state (EOS) of hot QCD is therefore of great
interest, also in order to reliably identify and calculate experimental signatures of that elusive
state. Perturbative results on the EOS are available up to order $\mathcal{O}(g_s^5)$ \cite{ZK95}.
However, for temperatures of interest the strong coupling constant is presumably large: $g_s \simeq 1 - 2$. The perturbative expansion in powers of $g_s$  shows bad convergence already for much smaller
values of the coupling. Furthermore, in the vicinity of a phase transition, perturbative methods are in
general not expected to be applicable. Non-perturbative methods such as lattice QCD calculations become
mandatory. From these numerical simulations the EOS of a pure gluon plasma is known to high accuracy
\cite{BE96}, and there are first estimates for the continuum EOS of systems including quarks,
albeit still with unphysically large masses \cite{FPL00}.

Various interpretations of the lattice data have been attempted, most prominently as the EOS of a gas
of quark and gluon quasiparticles. In a phenomenological framework, quarks and gluons are simply
treated as non-interacting, massive quasiparticles \cite{PKS00}. More recently, a quasiparticle
description of QCD thermodynamics has been derived in a more rigorous treatment using resummed hard thermal loop (HTL)
perturbation theory \cite{BI01}. Employing the full HTL spectral representions, the resulting EOS can
be matched to lattice data down to temperatures $T \sim 3 \ T_c$; below that temperature,
non-perturbative physics not amenable in an expansion in $g_s$ becomes important. Unfortunately, as
evident from figure \ref{figure1} (left panel), current experiments only probe that very temperature
regime where the underlying physics, the confinement and chiral symmetry breaking mechanism, is not
sufficiently understood. Phenomenological models incorporating as much physics as is known are
therefore necessary. Here, we propose a new quasiparticle model of the QGP that incorporates a parametrization
of confinement close to $T_c$, supplemented by thermal masses compatible with lattice results. Details
can be found in \cite{RAS}.
%
\section{Quasiparticles and confinement}
%
The use of a quasiparticle model in QCD is based on the observation
that in a strongly interacting system, the complex dynamics often rearranges itself in such
a way that gross features
of the physics can be described in terms of appropriate effective degrees of
freedom. Consider a SU(3) gluon plasma. From asymptotic freedom, we expect that at very high temperatures the plasma
consists of quasifree gluons. As long as the spectral function of the thermal
excitations at lower temperatures resembles qualitatively this asymptotic form, a
gluonic quasiparticle description is expected to be applicable. Their dispersion relation reads
\begin{equation}
\omega^2_k \simeq k^2 + m^2_*(T), \label{disp_rel}
\end{equation}
where $m_*(T)$ acts as an effective mass generated dynamically by the interaction of the gluons with
the heat bath background.
%
\begin{figure}[htb]
%
\begin{center}
\epsfig{file=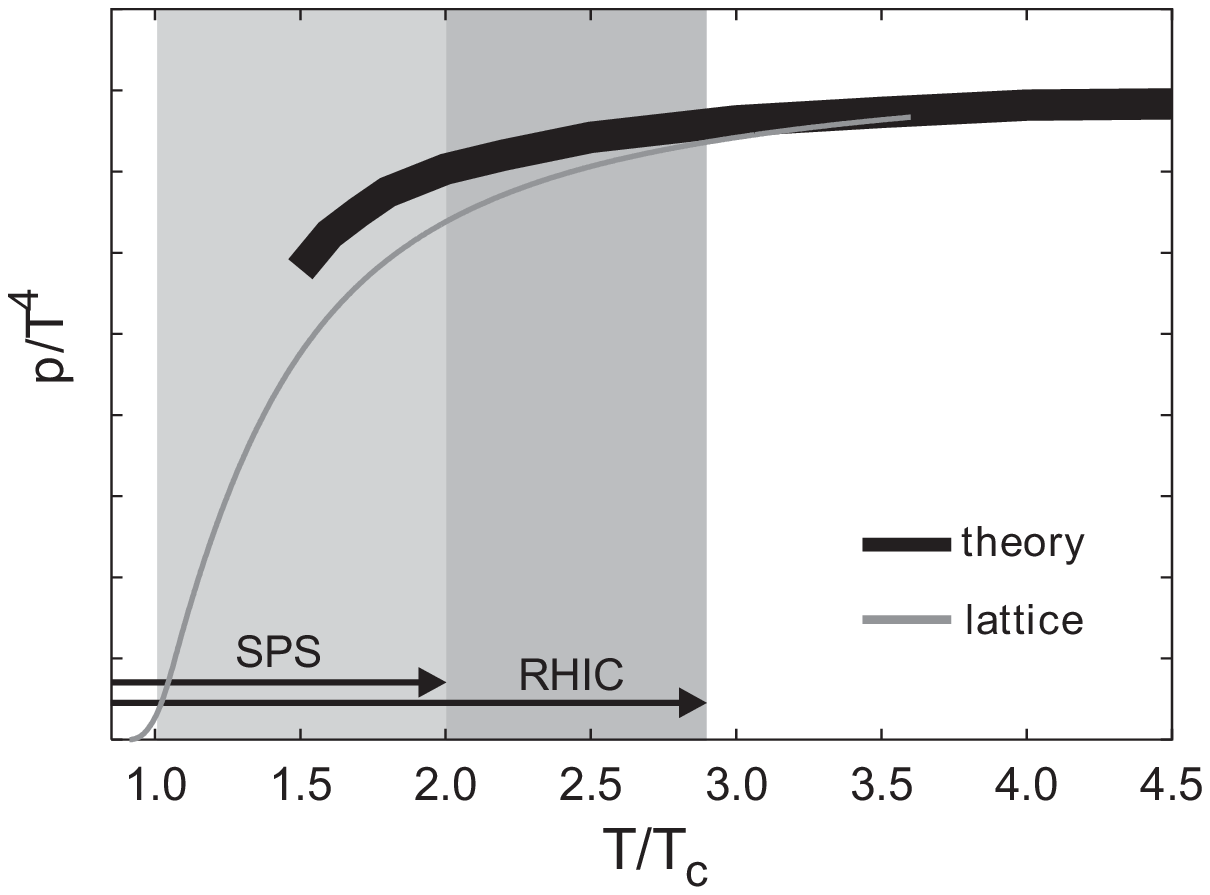,width=6cm} \epsfig{file=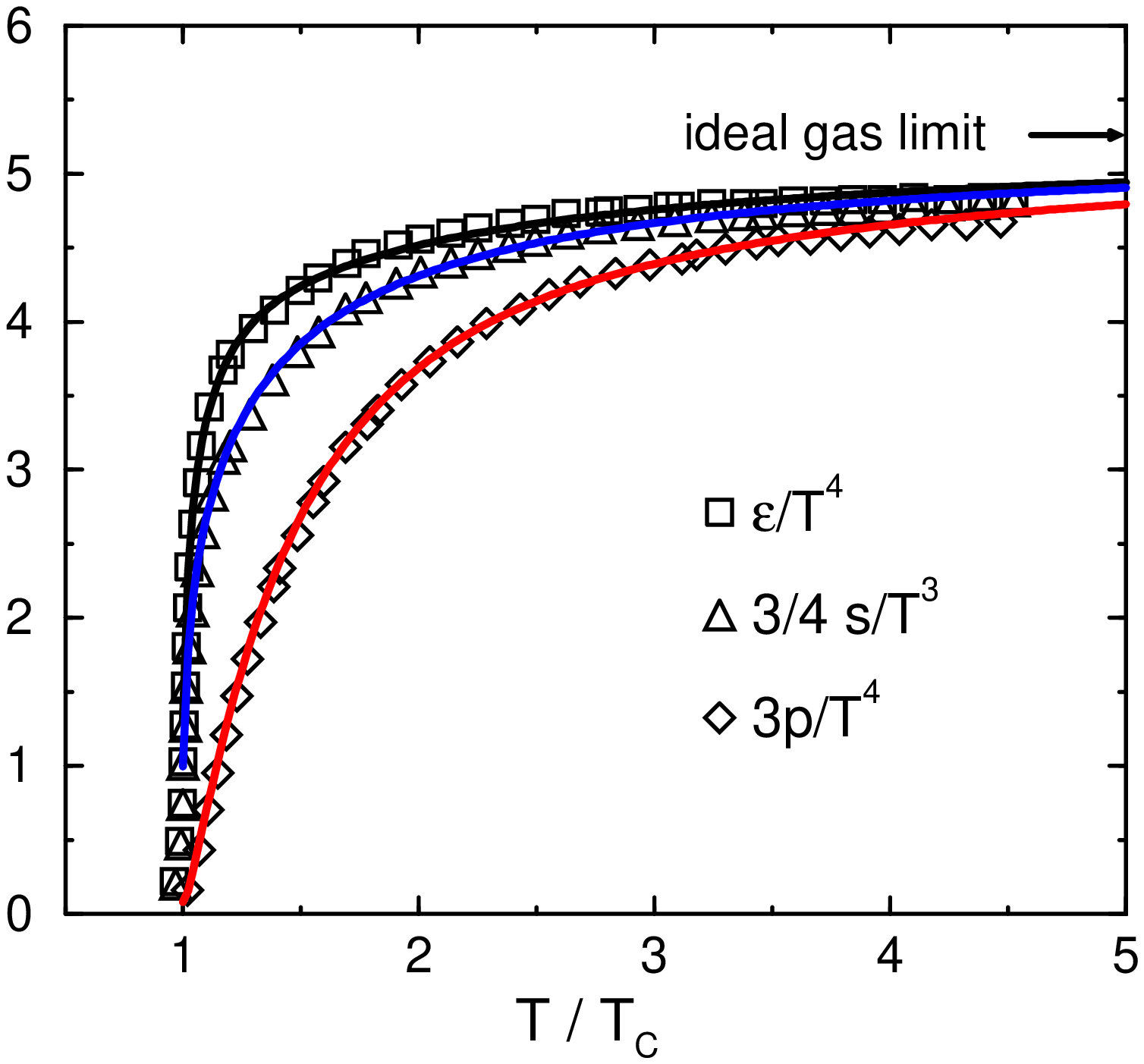,width=7cm}
\caption{Left panel: Sketch of theory vs. lattice and the temperatures probed by current experiments.
Right panel: Normalized $\epsilon$, $s$ and $p$ (solid lines) compared to continuum extrapolated SU(3)
lattice data (symbols) \cite{BE96}.} \label{figure1}
\end{center}
\vspace{-0.5cm}
\end{figure}

However, the picture of a simple massive gas is presumably not appropriate close to $T_c$ because the
driving force of the transition, the confinement process, is not taken into account. This physics has
to be inserted by hand. Below $T_c$, the relevant degrees of freedom in SU(3) gauge theory are heavy,
colour singlet glueballs. Approaching $T_c$, deconfinement sets in and the gluons are liberated,
followed by a sudden increase in entropy and energy density. Conversely, when approaching the phase
transition from above, the number of thermally active degrees of freedom is reduced due to the onset of
confinement. As $T$ comes closer to $T_c$, an increasing number of gluons gets trapped in glueballs
which {\em disappear} from the thermal spectrum: since $m_{GB} \sim 1.5 $ GeV and $T_c \sim 270$ MeV
(for pure gauge theory), glueballs are simply too heavy to become thermally excited in the temperature
range under consideration (up to about $5 \ T_c$). So, while the confinement mechanism as such is still
not understood, it is not necessary to know it in detail since we consider a {\em statistical} system.
All confinement does on a {\em large} scale is to cut down the number of thermally active gluons as the
temperature is lowered. This effect can be included in the quasiparticle picture by modifying the distribution function of the gluons by a temperature-dependent {\em confinement factor $C(T)$}: $f_B(\omega_k)
\rightarrow C(T) f_B(\omega_k)$.

To become quantitative, we have to specify the thermal masses $m_*(T)$ entering (\ref{disp_rel}). Based
on the observation that the Debye screening mass $m_D$ evaluated on the lattice shows approximate
critical behaviour \cite{KK00}, in accordance with a weakly first order phase transition and in
contrast to perturbative results, we parametrize
\begin{equation}
m_*(T) \sim G_0  T \left(1 -  \frac{T_c}{T} \right)^{\beta}. \label{m_g}
\end{equation}
Consider now the entropy of a gas of massive gluons with such a dropping $m_*(T)$. The result for
$s(T)$ will clearly overshoot the lattice entropy because light masses near $T_c$ lead to an increase
in $s(T)$. However, since the entropy is a measure for the number of active degrees of freedom, the
difference may be accounted for by the aforementioned confinement process as it develops when the
temperature is lowered towards $T_c$. The explicit temperature dependence of the confinement factor
$C(T)$ can be obtained simply as the ratio of the lattice entropy and the entropy calculated with a
dropping input gluon mass, and $C(T)$ again shows near-critical behaviour: $C(T) \sim (1 - T_c/T)^\gamma$.

\begin{figure}[bth]
%
\begin{center}
\epsfig{file=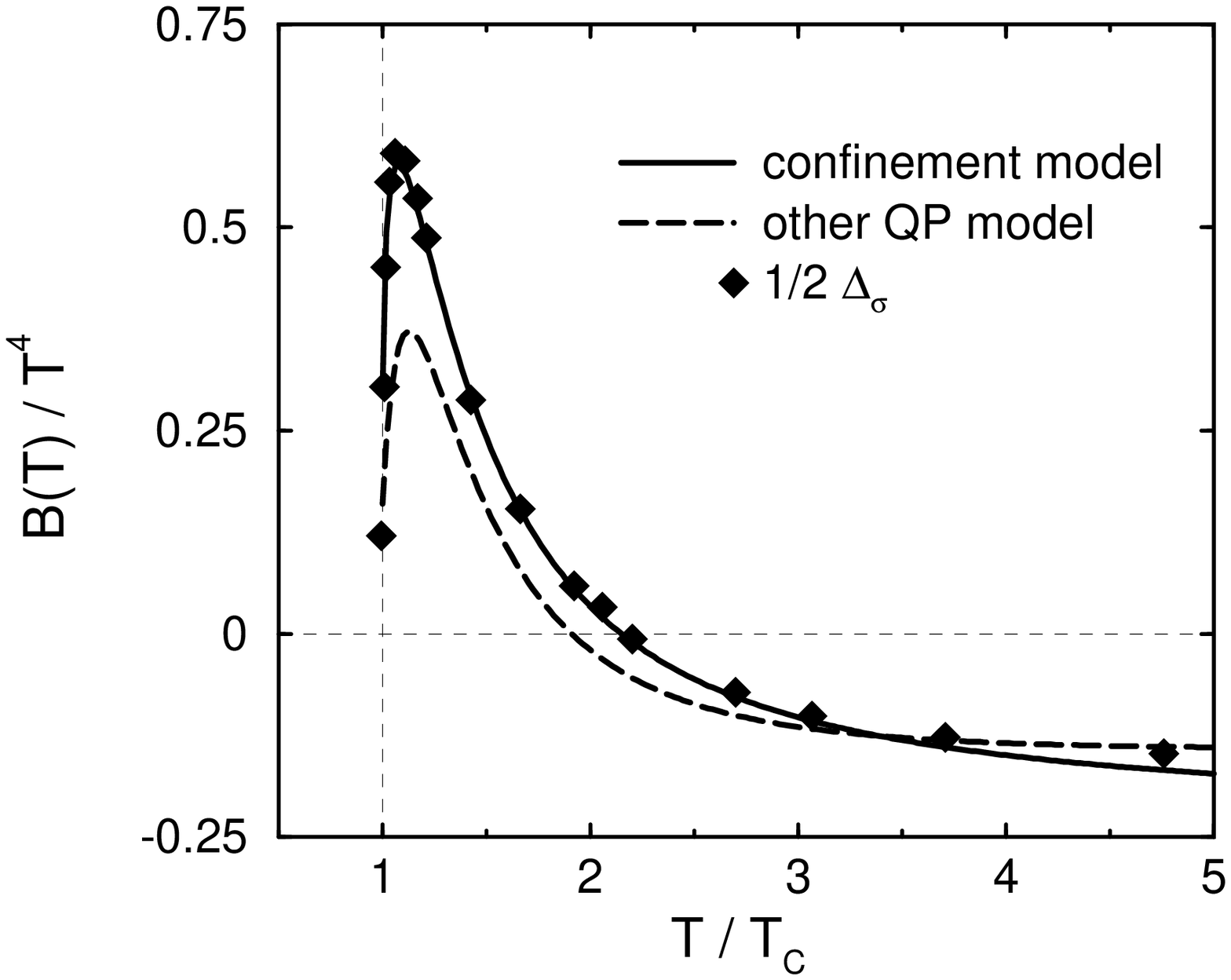,width=6.5cm} \epsfig{file=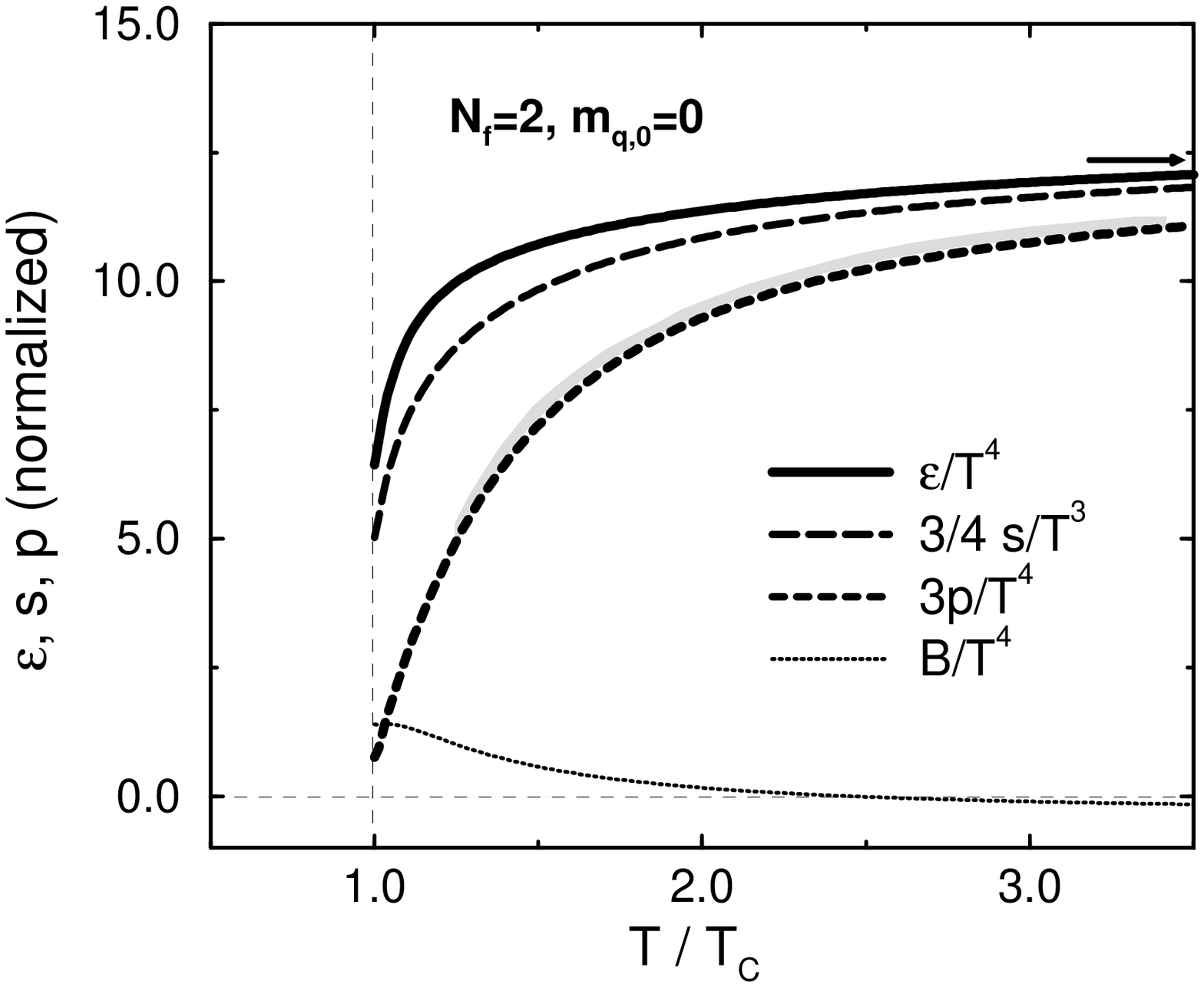,width=6.5cm}
\caption{Left panel: The function $B(T)$. Symbols show $\frac{1}{2}\Delta_\sigma$ \cite{BE96}, the
dashed line displays $B(T)$ in other quasiparticle models \cite{PKS00}. Right panel:  Normalized
$\epsilon$, $s$ and $p$ (solid lines) for $N_f = 2$ in the chiral limit. The grey band is a
corresponding lattice estimation \cite{FK02}.} \label{figure2}
\end{center}
\vspace{-0.5cm}
\end{figure}

\vspace{0.5cm}
\noindent
Thermodynamical quantities like the energy density can now be calculated as
\begin{equation}
\epsilon(T) = \frac{N_c^2 - 1}{2\pi^2} \int \limits_0^\infty dk \ k^2  \left[ C(T) f_B(\omega_k) \right] \
\omega_k + B(T).
\end{equation}
The function $B(T)$ is not an independent function, but uniquely determined by $m_*(T)$, $C(T)$ and
their $T$-derivatives. It is necessary to maintain thermodynamical self-consistency and can be
interpreted as the thermal energy density of the vacuum. The explicit expressions for $B(T)$, the
pressure and the entropy density can be found in \cite{RAS}. In figure \ref{figure1} (right panel), we
compare $\epsilon$, $s$ and $p$ with lattice data, as a function of $T$. We achieve a good and economic
parametrization. The so-called interaction measure, related to the trace of the energy-momentum tensor,
$T^\mu_{\ \mu} = \epsilon - 3p$, is also nicely reproduced. Figure \ref{figure2} (left panel) shows the
function $B(T)$ and the spacelike plaquette expectation value $\Delta_\sigma$, as measured on the
lattice, that can be related to the thermal chromomagnetic condensate $\langle \mathbf{B}^2 \rangle_T$.

%
%
%
\noindent We find the simple relation
$$
B(T) = \frac{1}{2}\Delta_\sigma(T) T^4 = - \frac{11 \alpha_s}{8\pi} \langle \mathbf{B}^2 \rangle_T +
\frac{1}{4} \langle G^2 \rangle_{T=0}
$$
with the zero-temperature condensate $\langle G^2 \rangle_{T=0}$. This relation between $B(T)$ and
$\langle \mathbf{B}^2 \rangle_T$ may hint at a deeper connection between $B(T)$ as a carrier of
non-perturbative effects, and the magnetic condensate. After all, $B(T)$ represents the thermal energy
of the (non-trivial) Yang-Mills vacuum. Note that there is no such relation in other quasiparticle
models, as evident from the dashed line in figure \ref{figure2} (left panel).
%
\section{Dynamical quarks}
%
The extension of the mechanism presented so far to systems with dynamical quarks is not
straightforward. Simulations of fermions on the lattice are still plagued by problems. However, when
plotting the lattice pressure, normalized to the ideal gas value, for the pure gauge system and for
systems with 2, 2+1 and 3 quark flavours, it is found that the QCD EoS shows a remarkable flavour {\em
independence} when plotted against $T/T_c$. The flavour dependence is then well approximated by a term
reminiscent of an ideal gas $p(T, N_f) \sim \left(16 + 10.5 \ N_f \right) \tilde{p}(T/T_c)$ with a
universal function $\tilde{p}(T/T_c)$ \cite{FK01}. This hints at a confinement mechanism being only
weakly flavour-dependent, and hence we assume that the function $C(T)$ acts in a universal way on
quarks and gluons. Figure \ref{figure2} (right panel) shows pressure, energy and entropy density for
two massless quark flavours. Reassuringly, the pressure of the confinement model lies well within a
narrow lattice estimate for the continuum EOS in the chiral limit \cite{FK02}. The picture for two
light and a heavier strange quark looks similar although the approach to the ideal gas limit is slower.

The neglect of the confined particles now violates detailed balance more severely since the lightest
Goldstone modes have masses comparable to $T_c$, hence they do contribute to the thermodynamics. Within
a statistical hadronization model it may be possible to estimate their contribution. Furthermore, the
calculation of susceptibilities will enable us to gain further insight in the nature of $C(T)$. Work
along these lines is in progress.
%
%
\section{Dilepton rates}
%
%
We apply the proposed model now to dilepton production in URHIC where it enters in two ways: the EOS
serves as input for the construction of the fireball that is discussed in \cite{RENK}. Second, the
dilepton emissivity of a static hot spot,
\begin{equation}
\frac{d\mathcal{N}_{ee}}{d^4x d\omega d^3 k} \sim \frac{\rho_V(\omega, k; T)}{\exp(\omega/T) -1},
\label{rate}
\end{equation}
is proportional to the vector spectral function $\rho_V$ that is calculated from the photon self
energy. By construction, the quasiparticle $q\bar{q}$-loop is the only contribution. In addition, the
confinement mechanism reduces the probability of thermally exciting a quark by $C(T)$ and thus the
total rate by $C(T)^2$. The corresponding $\rho_V$ agrees with qualitative lattice results
\cite{LAT_SPEC}.

To compare with experiment, the rate (\ref{rate}) has to be convoluted with the fireball expansion. A
discussion of the fireball dynamics and the final dilepton rates can be found in T. Renk's proceedings contribution
\cite{RENK}.
%
%
%

%
\end{document}